\documentclass[12pt]{article}

\usepackage{amssymb,amsmath,xcolor,graphicx,xspace,colortbl,ragged2e,rotating} %
\usepackage{amsfonts}  
\usepackage{amsmath}  
\usepackage{amssymb}  
\usepackage{amssymb,amsmath,xcolor,graphicx,xspace,colortbl,ragged2e,rotating}  
\usepackage{graphicx}  
\graphicspath{{name17_graphics/}{name17_tcache/}{name17_gcache/}}
\DeclareGraphicsExtensions{.pdf,.eps,.ps,.png,.jpg,.jpeg}
\graphicspath{{SO10_graphics/}{SO10_tcache/}{SO10_gcache/}}
\DeclareGraphicsExtensions{.pdf,.eps,.ps,.png,.jpg,.jpeg}
\providecommand{\U}[1]{\protect\rule{.1in}{.1in}}
\newtheorem{theorem}{Theorem}

\newtheorem{definition}[theorem]{Definition}

\begin{document}
\title{Particles as superoscillations of spacetime with a nonlocal metric?}
\author{Tomer Shushi  \\\relax Ben-Gurion
University of the Negev }
\maketitle
\title{}
\begin{abstract}Einstein field equations show how matter curve spacetime, but, does curved spacetime creates matter? And if so, can we have geometrical foundations
to every matter in the universe? In this note, we suggest an approach to derive non-general relativistic dynamics of particles as curvatures of spacetime
under the assumption of nonlocality. In particular, we examine the possibility that particles are obtained by superoscillatory functions of spacetime. By
introducing a metric that has an impact on every point in spacetime, we give a precondition for nonlocality under this ontic model. The model is deterministic
and contains a nonlocal hidden variable. This hidden variable is the mass density of the global metric. Due to the uncertainty principle, this hidden variable
is hidden in the sense that for getting full information about it one should concentrate energy/momentum in a small volume in spacetime that it will create
a black hole which will destroy the mass-density\ at that particular area. Therefore, it remains hidden by the
protection of spacetime itself and its geometric structure. 
\end{abstract}

\section{Introduction}
The heart of the geometrical approach to understanding quantum physics relies on the insight that Einstein's field equations directly imply that
\begin{equation}\left \{\text{Curvature of spacetime} \Longleftrightarrow \text{Matter}\right \} \label{1}
\end{equation}which means that as matter forms curvatures of\ spacetime,
curved spacetime also forms matter. This profound insight implies that matter can be originated from the geometry of spacetime. The problem of such equivalence
(\ref{1}) comes when we consider quantum particles. In that case, although it is straightforward that matter curves spacetime,
it is not trivial that curvature of spacetime creates matter. Therefore, many attempts were made to reconcile curvatures of spacetime with particle physics
[1-11]. 

The spacetime metric in general relativity is defined by a symmetric metric $g_{\mu  \nu }$ that describes the\ spacetime manifold. Many nonlocal hidden variables theories attempt to
describe the quantum world with an ontological interpretation of the wavefunction [13-17]. For example, in [17], a nonlocal global random variable is defined
in order to derive the non-relativistic spin-0 Schr{\"o}dinger equation. 

Let $\Omega _{\mu  \nu }$ be a spacetime metric, called the \textit{global metric} that is nonlocal with respect to $t =x_{0}$ and $\left (x_{1} ,x_{2} ,x_{3}\right ) ,\;$the time and the three spatial dimensions, respectively. It represents a small perturbation of spacetime (in
Planck scale).\ We assume that the global metric\ satisfies the following
properties: (P1) Global-nonlocality. We define the concept of nonlocality in the following way:\ \textit{Changing
the global metric in one region in spacetime would simultaneously change other separate regions}. (P2) The existence of a global mass density.
We assume the existence of a global mass density, $\rho _{\Omega } \left (t ,\mathbf{x}\right )$, associated with the spacetime curvature of $\Omega _{\mu  \nu }\text{,}$ that is conserved in the sense that $\int \rho _{\Omega } \left (t ,\mathbf{x}\right ) d \mathbf{x} =\lambda $ is a constant for each $t\text{.}$ (P3) The scalar curvature (Ricci scalar), $\mathcal{R}^{\Omega } \left (\rho _{\Omega }\right )$, of the global metric satisfies the homogeneity property $a \cdot \mathcal{R}^{\Omega } \left (a \cdot \rho _{\Omega }\right ) =\mathcal{R}^{\Omega } \left (\rho _{\Omega }\right )$ for any constant $a \in \mathbb{R}\text{.}$ Now, let $g_{\mu  \nu } \in \mathcal{G}$ where $\mathcal{G}$ is the set of spacetime pseudo-Riemannian metrics that creates very small perturbations of spacetime, i.e.,
$g_{\mu  \nu } =\eta _{\mu  \nu } +\epsilon _{\mu  \nu }$ where $\eta _{\mu  \nu }$ is the Minkowski metric with signature $( -1 ,1 ,1 ,1)$ and $\epsilon _{\mu  \nu }$ contains $\epsilon  -$small (almost vanished) components. Following our assumptions about the existence of the global metric $\Omega _{\mu  \nu }\text{,}$ in order to define an object with $g_{\mu  \nu }$ we will add it the global metric $\Omega _{\mu  \nu }\text{,}$ which will be denoted by $\mathcal{Z}_{\mu  \nu }\text{,}$
\begin{equation}\mathcal{Z}_{\mu  \nu } =g_{\mu  \nu } +\Omega _{\mu  \nu } . \label{m1}
\end{equation}Taking the summation of two metrics is the standard\ way
to describe\ the influence of two metrics in\ spacetime. For example,
it is common to define a metric that has a small spacetime perturbation by $\eta _{\mu  \nu } +h_{\mu  \nu }$ where $\eta _{\mu  \nu }$ is the flat metric (Minkowski metric), and $h_{\mu  \nu }\;$is some small deviation from $\eta _{\mu  \nu }\text{.}$ The above metric will play an essential role in the sequel, and as so, we are interested in its fundamental
properties. 

The metric of (\ref{m1}) is given by
\begin{align*}d s^{2} &  =\mathcal{Z}_{\mu  \nu } d x_{\mu } d x_{\nu } \\
 &  =d s_{g_{\mu  \nu }}^{2} +d s_{\Omega _{\mu  \nu }}^{2}\text{.}\end{align*}where $d s_{g_{\mu  \nu }}^{2} =g_{\mu  \nu } d x_{\mu } d x_{\nu }$ and $d s_{\Omega _{\mu  \nu }}^{2} =\Omega _{\mu  \nu } d x_{\mu } d x_{\nu }$ describe the spacetime curvature with respect to $g_{\mu  \nu }$ and $\Omega _{\mu  \nu }\text{,}$ respectively. 

The Christoffel symbols are then
\begin{equation}\Gamma _{\mu  \nu }^{\rho } =\frac{1}{2} \mathcal{Z}^{\rho  \alpha } \left (\frac{ \partial \mathcal{Z}_{\alpha  \mu }}{ \partial x^{\nu }} +\frac{ \partial \mathcal{Z}_{\alpha  \nu }}{ \partial x^{\mu }} +\frac{ \partial \mathcal{Z}_{\mu  \nu }}{ \partial x^{\alpha }}\right ) , \label{Cr1}
\end{equation}with the standard notions of the Ricci tensor and scalar curvature (Ricci scalar) that take the following
forms
\begin{equation*}\mathcal{R}_{\mu  \nu } = \partial _{\alpha }\Gamma _{\nu  \mu }^{\alpha } - \partial _{\nu }\Gamma _{\alpha  \mu }^{\alpha } +\Gamma _{\alpha  \rho }^{\alpha } \Gamma _{\nu  \mu }^{\rho } -\Gamma _{\nu  \rho }^{\alpha } \Gamma _{\alpha  \mu }^{\rho }\text{,}
\end{equation*}and $\mathcal{R} =\mathcal{Z}^{\mu  \nu } \mathcal{R}_{\mu  \nu }\text{,}$ respectively, and the Einstein-Hilbert (EH) action with respect to $\mathcal{Z}_{\mu  \nu }$ is given by $S \left [\mathcal{Z}_{\mu  \nu }\right ] =\frac{1}{16 \pi } \int \mathcal{R} \sqrt{ -\det  \left (\mathcal{Z}_{\mu  \nu }\right )} d t d^{3} \mathbf{x}\text{.}$ 

For the sequel, we define two types of the classical limit. We denote \textit{(relativistic)
classical limit} as the limit from a relativistic into a non-relativistic theory, and \textit{(quantum) classical limit} as the limit
from quantum mechanics to classical mechanics. 

\section{The Main Idea }
In Fourier analysis, it was believed\ that band-limited signals in space and time could not oscillate
locally arbitrarily faster than the highest component in their spectrum. However, it was discovered that there is a unique type of signals that violate
such a feature; these are the superoscillatory functions [18-30]. Superoscillations\ show that a superposition
of small Fourier components that have bounded Fourier spectrum has the possibility to have a shift that is well outside the spectrum. These functions appear
in both classical and quantum systems. For example, such functions have been observed and investigated in classical optics, image processing, matter waves,
electric and magnetic fields, and even in engineering [19,20,24,27-29]. 

Superoscillations are defined in the following way [20]:

\begin{definition}
 \bigskip A superposition function
\begin{equation*}\sum _{j =0}^{n}C_{j} \left (n ,q\right ) e^{i S_{n} (j) x} ,q ,\alpha  \in \mathbb{R}_{ +} ,x \in \mathbb{R}\text{,}
\end{equation*}is called a superoscillatory function if it satisfies the following requirements: 

(S1) $\left \vert S_{n} \left (k\right )\right \vert  <\alpha $ for all $n$ and $k \in \mathbb{N} \cup \{0\}\text{.}$ 

(S2) There exist a compact subset $C$ of $\mathbb{R}\text{,}$ on which $\varphi _{n}$ converges uniformly to $e^{i \mathfrak{S} (q) x}$ where $\mathfrak{S}$ is a continuous real-valued function such that $\vert \mathfrak{S} (q)\vert  >\alpha \text{.}$ 
\end{definition}

In their seminal paper, Aharonov et al. (1990) [31] have shown
that suitable superposition of time evolutions due to the action of weak forces is equivalent to a time evolution resulting from the action of a strong
force, with a new action functional. This startling result is the heart of our work. In another work, Aharonov et al. (2018) [32] present a novel renormalization
technique based on a superposition of Lagrangians, which leads to a new form of Lagrangian that is renormalized. 

As defined earlier,
the surprising nature of superoscillations shows that band-limited functions can have oscillations arbitrarily faster than the fastest Fourier component
of the function. However, the inverse is also true, and provide a startling fact: A wave with arbitrary large oscillation can be described by a superposition
of waves with much lower oscillations that are independent on both the phase and the amplitude of the original wave. 

Our goal is to
model non-GR particles, i.e., particles in the microscopic scale\ that their dynamics do not follow the GR\ .
We suggest using superoscillations to obtain it. 

Let us consider a wave function with some dynamics, given by, \ $\Psi _{\mathcal{Z}} =A (x_{\mu }) e^{i \mathcal{S} [x_{\mu }]/h}\text{,}$ where $h \in \mathbb{R}$ is some real-valued variable that, in the sequel, will be fixed to a constant. This wave function
is not restricted to any specific dynamics. However, we define it in accordance with GR. (a) We assume that it represents the dynamics of the system with
symmetric time. (b) $\Psi _{\mathcal{Z}}$ is a continuous function. We also consider $\vert \mathcal{S} [x_{\mu }]\vert  >\alpha $. Then, without loss of generality, using superoscillations, for a constant $1/\vert h\vert  <m$ (for suitable $m >0$) one can convert $\Psi _{\mathcal{Z}}$ into a superposition of gravitational waves with bounded wavenumbers and specific amplitudes that change in $x_{\mu }$
\begin{equation}\Psi _{\mathcal{Z}} =A \left (x_{\mu }\right ) e^{i \mathcal{S} \left [x_{\mu }\right ]/h} \approx \sum _{j =0}^{n}\mathcal{A}_{j}^{\rho _{\Omega }} \left (x_{\mu } ;n\right ) e^{i k_{j}^{\mu } x_{\mu }/h} =\sum _{j =0}^{n}\varphi _{\mathcal{Z}_{j}} (n)
\end{equation}where each $\varphi _{\mathcal{Z}}$ represents a gravitational wave with amplitude $\mathcal{A}_{j}^{\rho _{\Omega }} \left (x_{\mu } ;n\right )\text{.}$ Since the phase of $\varphi _{\mathcal{Z}_{j}} (n)$ does not depend on $A$ nor $\mathcal{S}\text{,}$ it is clear that the phase and amplitude of $\Psi _{\mathcal{Z}}$ are encoded in $A_{j}^{\rho _{\Omega }} \left (x_{\mu } ;n\right )$. $k_{j}^{\mu }$ is a suitable wave vector of the j-th gravitational wave, for $x_{\mu }$ such that $\vert k_{j}^{\mu } x_{\mu }\vert  <\alpha $ . 

When taking the limit $n \rightarrow \infty $, we have that
\begin{equation}\vert \Psi _{\mathcal{Z}} -\sum _{j =0}^{n}\varphi _{\mathcal{Z}_{j}} (n)\vert  \rightarrow 0\text{    as    }n \rightarrow \infty .
\end{equation}We conclude\textit{ out-of-the-spectra} dynamical systems where the evolution is outside the scope of the GR\ evolutions,
seemingly showing a new type of dynamics that evolve from the GR one. At first glance, superoscillations are seemed to be a very unique form of waves, and
thus, one may assume that they should be very rare in nature. However, it turns out that even for generic random fields (e.g., Gaussian-like random fields)
superoscillations [38] may exist in about 1/3 fraction area of the random field. Thus, it is possible that superoscillations of spacetime can be generated
frequently in nature, and do not need special environment in order to evolve. 

The Madelung equations provide a hydrodynamic interpretation
of quantum mechanics [33-37].\ These coupled differential equations are derived from the Schr{\"o}dinger equation,
where the first one describes a classical continuity equation, while the latter equation describes the classical Euler equation with the addition of a nonlocal
potential, called the quantum potential [35]. In the Madelung equations, the quantum potential is given by
\begin{equation}Q = -\frac{1}{2 m} \frac{ \nabla ^{2}\sqrt{\rho }}{\sqrt{\rho }} , \label{Q1}
\end{equation}where $\rho $ is the mass density of the quantum wavefunction $\psi  =\sqrt{\rho } e^{i \mathcal{S}/h}$ in the polar representation, with the units of Planck constant $h =\hbar  =1\text{,}$ and $m$ is the mass of the particle. The quantum potential\ (\ref{Q1})
depends on the curvature of the density of the quantum particle, and it gives a precondition for the existence of nonlocality, since\ multiplying
the wave function by some constant (which changes its magnitude) will not change the quantum metric\ (see, for
instance,\ [35]). 

We now show how the global metric provides a precondition for nonlocality.

This will be represented by the Ricci scalar for each $j$-th metric $\mathcal{Z}_{\mu  \nu } (j)$. One can observed that $\mathcal{R}^{\mathcal{Z}} (j ;\rho _{\Omega })$ can be decomposed to the sum of Ricci scalar of $\Omega _{\mu  \nu }$ and other scalar curvature that depends on both\ $\Omega _{\mu  \nu }$ and $g_{\mu  \nu } (j) ,j =1 ,2 ,\ldots  ,n\text{,}$
\begin{equation}\mathcal{R}^{\mathcal{Z}} (j ;\rho _{\Omega }) =\mathcal{R}^{\Omega } \left (\rho _{\Omega }\right ) +\mathcal{U}_{\mathcal{Z}} (j ;\rho _{\Omega }) , \label{T1}
\end{equation}where $\mathcal{R}^{\Omega } \left (\rho _{\Omega }\right )$ is the Ricci scalar of the global metric, and $\mathcal{U}_{\mathcal{Z}} (j ;\rho _{\Omega })\text{.}$ 

Using (\ref{T1}) and from the property (P3) of the global metric,
we clearly see that for any real constant $a\text{,}$
\begin{equation}a \cdot \rho _{\Omega } \mathcal{R}^{\mathcal{Z}} (j ;a \cdot \rho _{\Omega }) =\rho _{\Omega } \mathcal{R}^{\Omega } \left (\rho _{\Omega }\right ) +a \cdot \rho _{\Omega } \mathcal{U}_{\mathcal{Z}} (j ;a \cdot \rho _{\Omega }) . \label{F1}
\end{equation}The precondition for nonlocality of $\mathcal{R}^{\Omega } \left (\rho _{\Omega }\right )$ comes from the fact that changing the magnitude of the mass density $\rho _{\Omega }$ by some constant $a \in \mathbb{R}$ will not change the scalar curvature of $\mathcal{R}^{\Omega } \left (\rho _{\Omega }\right )\text{.}$ Therefore, $\mathcal{R}^{\Omega } \left (\rho _{\Omega }\right )$ is independent on the\ field intensity of $\Omega _{\mu  \nu }\text{.}$ Thus, the Ricci scalar of the $j$-th state is the sum of local scalar $\mathcal{U}_{\mathcal{Z}} (j ;\rho _{\Omega })$ and a nonlocal scalar curvature $\mathcal{R}^{\Omega } \left (\rho _{\Omega }\right )\text{.}$ 

The theory contains a hidden variable, which is the mass density of the global
metric.\ Due to the uncertainty principle, the mass density of the metric in a Planck volume\ is
hidden in the sense that for getting full information about it one should concentrate energy/momentum in such small volume in spacetime, which\ will
create a black hole that will destroy such a metric at that particular area. Thus, this hidden variable remains\ hidden
by the protection of spacetime itself and its geometrical structure. An implication of\ nonlocality in a relativistic
causal model is that it evolves the uncertainty principle. This comes from the fact that\ nonlocality and relativistic
causality give the uncertainty principle, and thus, every relativistic\ nonlocal model contains the uncertainty
principle [39]. The classical limit is obtained due to the mass-density of an object that creates its own curvature, that is much larger than the one created
by $\Omega _{\mu  \nu }$ which becomes negligible including its nonlocality, and the uncertainty principle disappears. In string theory, the spin of the particles
comes from the rotation of the string that represents the particle. In our model, the particle is an oscillation of spacetime that is built from spacetime
metrics, which allows, conceptually, to understand the spin of the particles by inner rotations of these sub-metrics $\mathcal{Z}_{\mu  \nu } (j)$. 

\textbf{The Global metric as a conformal transformation of flat spacetime.} From properties (P1)-(P3) we
observe that $\Omega _{\mu  \nu }$ depends on the\ global mass density, and it satisfies the homogeneity property shown in (P3).
In the paper of Delphenich [34], it was proved that by taking a conformal transformation of the Minkowski metric $\eta _{\mu  \nu }$ we can build a nonlocal quantum potential. We now use this example in favor of our model. Suppose that the global metric is given by
a conformal transformation of the flat spacetime $\Omega _{\mu  \nu } =\eta _{\mu  \nu } e^{2 \sigma }\text{,}$ where $e^{2 \sigma } = :\rho _{\Omega }\;$is the global mass density. Moreover, assume that each amplitude of the $j -$th state $\varphi _{\mathcal{Z}}^{(j)}$\ is given by\ $\mathcal{A}_{j} \left (t ,\mathbf{x}\right ) =\sqrt{\rho _{\Omega }} \cdot a_{j} \left (t ,\mathbf{x}\right )$ where $a_{j} \left (t ,\mathbf{x}\right )$ is some positive and smooth function. Then, a straightforward calculation of the scalar curvature yields to
\begin{equation}\rho _{\Omega } \mathcal{R}^{\mathcal{Z}} (j ;\rho _{\Omega }) = -\frac{\square \sqrt{\rho _{\Omega }}}{\sqrt{\rho _{\Omega }}} +\rho _{\Omega } \mathcal{U}_{\mathcal{Z}} (j ;\rho _{\Omega }) , \label{rhor1}
\end{equation}where $\square  = - \partial _{t}^{2} +  \partial _{x_{1}}^{2} + \partial _{x_{2}}^{2} + \partial _{x_{3}}^{2}$ is the d'alembert operator.\ In fact, eq. (\ref{F1}) shows
that the Ricci scalar, $\mathcal{R}^{\mathcal{Z}}\text{,}$\ coupled to the mass density, $\rho _{\Omega }\text{,}$\ have a contribution that does represent a "quantum correction"
to the model (see, again, [34]). 

\textbf{The nature of the global metric hidden variable.} Since the curvature of the global metric\ is
very small, in particular, it is a Planck length perturbation of spacetime.\ Due to the uncertainty principle,
when we measure such a metric we must concentrate high energy/momentum in such a small volume in order to get information about the metric, a black hole
immediately emerges and destroy this particular area. Therefore, one physically cannot know the exact behavior of the particle, which is defined explicitly
by $\varphi _{\mathcal{Z}_{j}}\text{.}$ 

\section{Discussion}
The quest for geometrical foundations to non-GR dynamics has been extensively studied since Einstein's attempt to reconcile Maxwell's equations
and the GR field equations. We have examined the possibility that non-GR dynamics emerge from superoscillations of spacetime. This leads to the main features
of the theory: (a) Superoscillation of weak gravitational forces is equivalent to a non-GR force, with a new action different from the EH action. (b) The
theory contains a nonlocal\ hidden variable that is represented by a density function of the global metric.

\end{document}